# Method of increasing the information capacity of associative memory of oscillator neural networks using high-order synchronization effect

**Andrei Velichko**[*], **Maksim Belyaev, Vadim Putrolaynen, and Petr Boriskov**

Petrozavodsk State University, Petrozavodsk, 185910, Russia
* Corresponding author: e-mail velichko@petrsu.ru

**Abstract**

Computational modelling of two- and three-oscillator schemes with thermally coupled $VO_2$-switches is used to demonstrate a novel method of pattern storage and recognition in an impulse oscillator neural network (ONN) based on the high-order synchronization effect. The method ensures high information capacity of associative memory, i.e. a large number of synchronous states $N_s$. Each state in the system is characterized by the synchronization order determined as the ratio of harmonics number at the common synchronization frequency. The modelling demonstrates attainment of $N_s$ of several orders both for a three-oscillator scheme $N_s \sim 650$ and for a two-oscillator scheme $N_s \sim 260$.

A number of regularities are obtained, in particular, an optimal strength of oscillator coupling is revealed when $N_s$ has a maximum. A general tendency toward information capacity decrease is shown when the coupling strength and switch inner noise amplitude increase. An algorithm of pattern storage and test vector recognition is suggested. It is also shown that the coordinate number in each vector should be one less than the switch number to reduce recognition ambiguity.

The demonstrated method of associative memory realization is a general one and it may be applied in ONNs with various mechanisms and oscillator coupling topology.

## 1. Introduction

Usage of artificial neural networks [1] for information processing allows mastering the problems that arise when traditional computation schemes are applied in such areas as pattern and speech recognition [2]. Therefore the important research trends include studying the modes of oscillator neural networks (ONN) operation and training, implementation of associative memory modes based, for example, on weakly coupled phase oscillators (Kuramoto model) [3] or impulse oscillators [4]. The effect of synchronization plays a crucial role in ONN operation and is often used as a marker of ONN action, for example, in pattern recognition event.

In ONN the system demonstrates frequency and phase synchronization [5] – [8] and also synchronization of high order [9] – [11] at certain control parameters such as parameters of an oscillator scheme or coupling strengths between the oscillators. The proposed here method of pattern storage and recognition is based on high-order synchronization. In a lot of studies [8], [12], [13] patterns to be stored are expressed through a set of vectors. Vector coordinates contain information about the pattern and unambiguously associate it with one of possible variants.

There are some methods of associative memory realization based on oscillator elements synchronization in ONN, one of them is presented in paper [5]. To store **E** vector a phase-shift keying method of encoding is used in a fully connected massive of $N$ oscillators and the vector itself is specified by a certain combination of phase shift **E**=($\delta\varphi_1$, $\delta\varphi_2$…$\delta\varphi_N$) using a weight matrix according to Habbian rule. Such method of (phase) encoding enables storage of more than one vector at the same weight matrix, i.e. there are several stable phase states of a system. Maximal number of stored vectors will be determined by the number of oscillators $N$. The recognition procedure has two stages, at the first one a test vector **T** is specified by weight matrix setting, at the second stage the weights are sharply changed to the initial values (corresponding to the stored vectors) and the system arrives at one of stable combination of phase shift **E**. However, this phase method has the following drawbacks: $N^2$ couplings with tunable weights and a two-stage procedure of pattern recognition.

A second known method of associative memory realization is a frequency-shift keying method of encoding based on synchronized frequency shift [8]. According to this method vector **E** is stored through oscillator frequencies shift against the central frequency of oscillator array $F^0$ synchronization (on the first harmonic) per the values corresponding to the vector coordinates **E**=($\delta\omega_1$, $\delta\omega_2$…$\delta\omega_N$). Recognition of test vector **T** occurs at the reverse shift of frequencies and in case when the vectors coincide **T≈E** the synchronization indicating the fact of pattern recognition takes place**.** This method allows usage of an oscillator star configuration and only $N$ couplings, however, the disadvantage of this method is that just one vector is stored.

The present paper suggests a breakthrough method of associative memory realization when the array of coupled oscillators allows storage and recognition of variety of patterns and their number (information capacity of storage) depends on the oscillator number $N$, range of control parameters, network topology, strength of coupling between oscillators and noise level in the system. This result is achieved because we use high order synchronization effect in our ONN that possesses plenty of oscillator synchronous states defined as harmonic number ratio at common frequency of synchronization.

## 2 Method
### 2.1 General principle

Oscillator neural network is a system of $N$ coupled oscillators, they may be connected via electric (by resistors and capacitors) [14], [15], thermal [10], [11] and optic [16] couplings depending on the physical mechanism of oscillator interaction. In the general case there is a matrix of coupling strengths $\Delta_{i,j}$ (weights), where $i$, $j$ are the numbers of interacting oscillators and $\Delta_{i,j}$ denotes the value of the $i$-th oscillator effect on the $j$-th one. Oscillator network may form various topologies: fully connected – all-to-all; and not fully connected – bus, star, and ring. Fig. 1a-c show examples of two and three oscillators connections using topologies "star", "all-to-all" and an example of $N$ oscillators connection using a mixed topology (Fig. 1d)

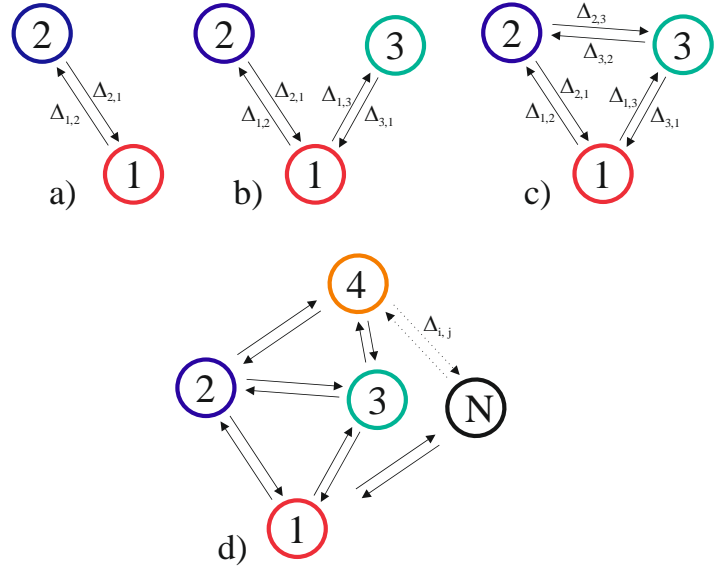

Fig. 1. Examples of two (a) and three (b) oscillators connection into a neural network using topologies "star" (b), "all-to-all" (c), and $N$-oscillators using mixed topology (d), where $\Delta_{i,j}$ indicates the value of the $i$-th oscillator effect on the $j$-th one.

It is known [9] – [11] that oscillators in a network may undergo the effect of synchronization and besides synchronization on the first harmonic synchronous modes of high order may be observed if signal spectra possess several harmonics. In the general case high-order synchronization is determined by the ratio $k_1:k_2:k_3:..k_N$, where $k_N$ is harmonic order of $N$-th oscillator at the common frequency of the network synchronization $F_s$. As an example Fig 2a shows spectra of three electric oscillators that have synchronization of the order $k_1:k_2:k_3=3:6:4$. The following rule should be noted: if all paired oscillators have different synchronization frequencies there is always a common synchronization frequency $F_s$ for the whole system (all pairs) and the network synchronous state will also be determined by the ratio $k_1:k_2:k_3:..k_N$ at frequency $F_s$ (see section 2.2)

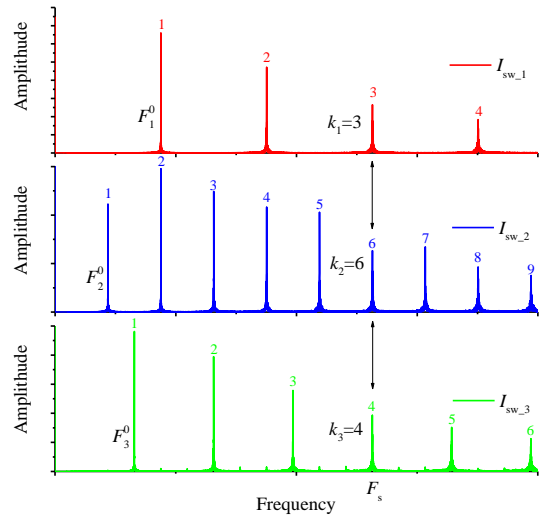

(a)

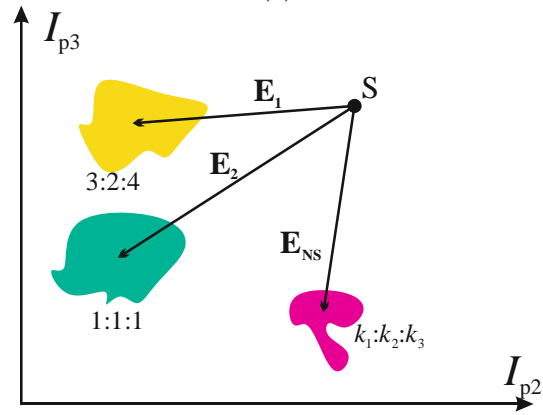

(b)

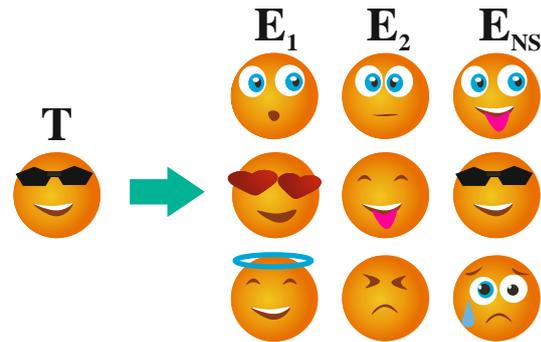

(c)

Fig. 2. a) Example of oscillation spectra of three electric oscillators at synchronization order $k_1:k_2:k_3=3:6:4$, b) schematic illustration of the synchronization areas for a three-oscillator scheme, c) example of vector and image association, where $I_p$ is current of a power source, $F^0$ is first harmonic, $k$ is the harmonic number at the synchronization frequency $F_s$.

Transition from one synchronous state into another is possible when the oscillator network control parameters are varied. For example, in electric oscillators the main parameters may be oscillator feed currents $I_p$, their variation causes changes of the basic oscillation frequency $F^0$. Nevertheless, in some cases, transition between states may be achieved by variation of coupling forces or noise intensity.

The range of control parameters variation where synchronization does not change its state is called a synchronization area. There is a whole family of synchronization areas that are called Arnold's tongues (for the case of two oscillators). A schematic example of synchronization areas for a three-oscillator scheme is shown in Fig. 2b. Here the control parameters are oscillator feed currents.

The number of possible variants of synchronous states (synchronization areas) where the system may exist when the basic control parameters are varied we shall call information capacity of associative memory $N_s$. The value of $N_s$ depends on the oscillator number $N$, control parameters range, network topology, strength of coupling between oscillators and noise level in the system. We will cover the issue in detail later, nevertheless, we have shown in [10] that for a two-oscillator network $N_s$ has a maximum at certain values of coupling strengths between oscillators and decreases when the system noise amplitude increases. When the coupling strength grows considerably the value $N_s$ decreases because of the nearby synchronization areas integration.

As we have mentioned in the introduction section, the patterns to be stored are usually expressed through a set of vectors. Vector coordinates contain information about the pattern and unambiguously associate it with one of possible variants (see example in Fig. 2c).

We suggest the following method of associative memory realization in a neural network based on high-order synchronization effect.

First, it is necessary to set the neural network at such initial point of control parameters ($S$) that is outside the system's synchronization areas (see Fig. 2b). While in a non-synchronous state the system may potentially reach one of various synchronization states ($N_s$) along the corresponding vectors E. Thus the system stores $N_s$ vectors at any point outside the synchronous areas, we shall denote them as $\mathbf{E}_1$, $\mathbf{E}_2$...$\mathbf{E}_{NS}$. The dimension of the stored vectors in this case is equal to the oscillator number $N$ and their coordinates determine the shift of oscillator control parameters, for instance, currents $\mathbf{E}_1=(\delta I_{p1}, \delta I_{p2}...\delta I_{pN})$, against the initial point S. Thus to store the certain vectors it is necessary to set the system initial point S in such non-synchronous state that the distances from this point to the existing synchronous states corresponded the stored vectors $\mathbf{E}$.

Second, when the initial point is set, to recognize the test vector $\mathbf{T}$ it is necessary to apply the shifts equal to the vector's coordinates to the control parameters. If one of the conditions is met ($\mathbf{T}=\mathbf{E}_1$ or $\mathbf{T}=\mathbf{E}_2$... or $\mathbf{T}=\mathbf{E}_{NS}$), i.e. coordinates values of $\mathbf{T}$ are equal to one of the stored patterns, a transition to the synchronous state will occur and, actually, the act of the corresponding pattern recognition will take place. Existence of synchronization areas ensures pattern recognition even at insignificant deviation of its coordinates from the stored pattern.

This method is a more complicated version of the method described in [8] where the analogy to the frequency-shift keying method of coding is used, just here the vector is specified through the control parameters $\mathbf{E}_1=(\delta I_{p1}, \delta I_{p2}…\delta I_{pN})$ instead of frequencies $\mathbf{E}=(\delta\omega_1, \delta\omega_2…\delta\omega_N)$, but the meaning is the same. The main difference of our method is in application of high-order synchronization effect that enables us to increase information capacity of associative memory of a neural network by many times.

Besides, as described in the results section, it is more practical to use vector dimensions $\mathbf{E'}=(\delta I_1, \delta I_2…\delta I_{N-1})$ with one less than the number of oscillators ($N$-1).

### 2.2 Model object

As a model object we have chosen a neural network composed of three thermally coupled $VO_2$-oscillators where each oscillator has the scheme of a relaxation oscillator. Our choice is conditioned by the fact that we have done some research in thermal coupling [11], [17] and its modeling, however, the coupling may be an electric one (capacitive or resistive [15]). It is known that an electric switching effect is observed in $VO_2$ film based structures that is conditioned by a phase metal-insulator transition (MIT) at the moment when the temperature reaches $T_t$ ~340 K because of Joule heating by the passing current $I_{sw}$ [18]. This gives high-impedance (OFF) and low-impedance (ON) branches on I-V characteristics with threshold voltages (OFF→ON) $U_{th}$ ~ 5 V and holding voltages (ON→OFF) $U_h$ ~ 1.5 V (see Fig. 3a). Both branches of I-V characteristics are reasonably well approximated by $f_{sw}$ curve consisting of two linearized regions with dynamic resistance $R_{off}$ ~9.1 kΩ и $R_{on}$ ~615 Ω:

$$I_{sw} = f_{sw}(U) \approx \begin{cases} \dfrac{U}{R_{off}}, & State=OFF \\ \dfrac{(U-U_{bv})}{R_{on}}, & State=ON \end{cases}, \quad (1)$$

where $U_{bv}$ ~ 0.82 V is bias voltage of a low-impedance region, and *State* is a switch state.

One of three presented in Fig. 1 topologies may be realized depending on coupling strength magnitudes Δ. At non-zero Δ≠0 the topology is "all-to-all" (Fig. 1c), at $\Delta_{2,3}=\Delta_{3,2}=0$ the topology is "star" (Fig. 1b), and at $\Delta_{2,3}=\Delta_{3,2}=\Delta_{1,3}=\Delta_{3,1}=0$ the scheme turns into a two-oscillator one (Fig. 1a). The control parameters here are source currents $I_{p1}$, $I_{p2}$, $I_{p3}$, their variation leads to alteration of the fundamental oscillation frequency $F^0$ of oscillators.

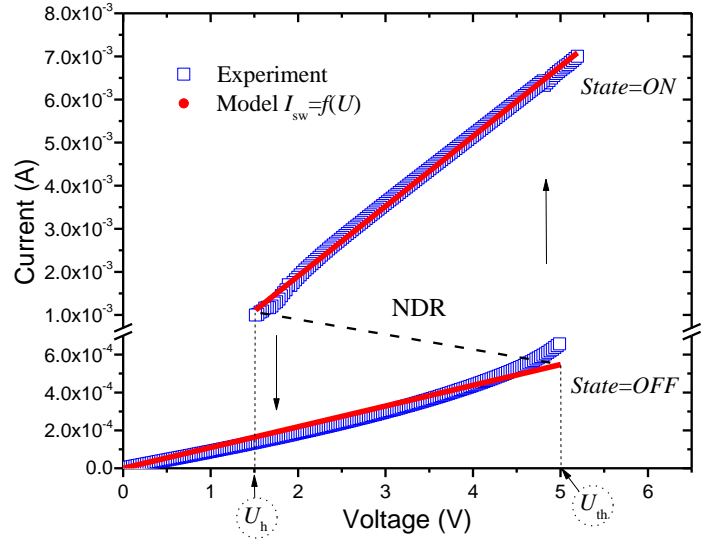

(a)

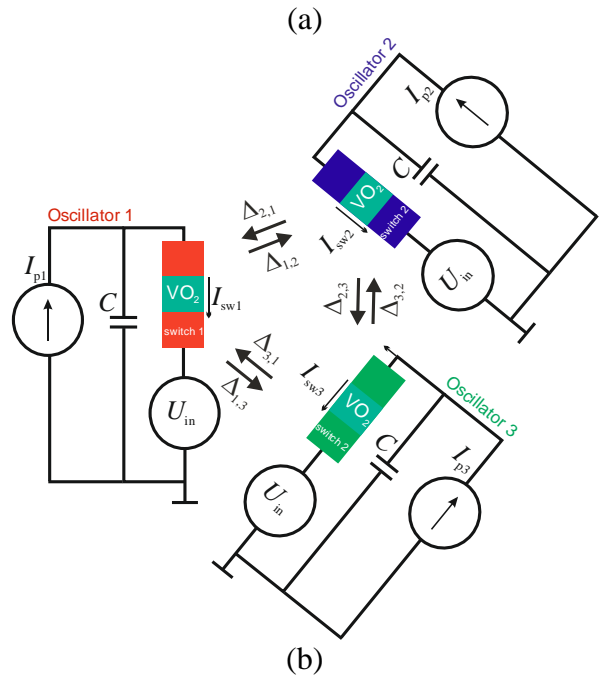

(b)

Fig. 3. Experimental and model I-V characteristics of VO$_2$-switch (a); a model scheme of a neural network based on three oscillators circuits with VO$_2$-switches interacting via thermal coupling (b).

Variations of each oscillator are described by the equation of Kirchoff's law:

$$C\frac{dU_i(t)}{dt} = I_{p(i)} - I_{sw(i)}(t), \qquad (2)$$

where $U_i(t)$ is output voltage taken from the capacitor ($C=100$ nF), $I_{sw(i)}(t)=f_{sw}(U_i(t)-U_{in})$ is current passing through a switch, determined by I-V characteristics (1), $I_{p(i)}$ is i-th oscillator supply current, respectively, $U_{in}$ is amplitude of switch internal noise, $i$ is oscillator's number.

Thermal interaction between the $i$-th VO$_2$-oscillator and the neighbor ones ( $(i+)$ – clockwise and $(i-)$ – counterclockwise of the scheme in Fig. 3b) is realized according to the rule:

$$U_{th(i)} = \begin{cases} U_{th(i)} - \Delta_{(i+),i}, & \text{if } State_{(i+)} = ON \\ U_{th(i)} - \Delta_{(i-),i}, & \text{if } State_{(i-)} = ON \\ U_{th(i)} - \Delta_{(i-),i} - \Delta_{(i+),i}, & \text{if } State_{(i-)} = State_{(i+)} = ON \end{cases} \quad . \quad (3)$$

If the states of oscillators $State_{(i+)}$ and $State_{(i-)}$ are on the OFF branch of I-V characteristics then the threshold voltage of the $i$-th VO$_2$-oscillator does not change: $U_{th(i)} = U_{th}$. Rule (3) is the same for all oscillators (with regard to cyclic permutation).

Oscillograms of oscillations with ~250 000 points and time interval $\delta t=10\mu s$ were simulated using (1-3). After that the oscillograms were automatically processed, synchronization order was determined and cross-sections of oscillator synchronization areas were built.

The switch parameters did not change in numerical simulation of the results, but current intensities $I_p$, coupling strength $\Delta$ and noise amplitude $U_{in}$ varied.

### 2.3 Method of synchronization order definition

The problem of finding the high-order synchronization value determined by the ratio of integers $k_1:k_2:k_3:..k_N$ (see section 2.1) may be solved in several ways. For example, by direct analysis of all oscillations spectra or by searching the synchronization order of each pair of oscillators based on the method which we suggested in [11].

It should be noted that at synchronous state the frequency sets of fundamental (first) harmonics of oscillators ($F_1^0, F_2^0, F_3^0,... F_N^0$) must be commensurable. This is evident, because at the synchronous state there is a common synchronization frequency $F_s$ and the equality ($F_s = F_1^0 \cdot k_1 = F_2^0 \cdot k_2 ... = F_N^0 \cdot k_N$) is fulfilled. If we divide $F_1^0$ into all frequencies in the set ($F_1^0, F_2^0, F_3^0,... F_N^0$), then we will get (1, $F_1^0/F_2^0, F_1^0/F_3^0,... F_1^0/F_N^0$)= (1, $k_2/k_1$, $k_3/k_1,...k_N/k_1$), that is a new set of rational numbers determining pair synchronization of all oscillators in regard to the first oscillator (see [11]).

Thus the method of specifying all values of $k$ and the synchronization order of the system consisting of $N$-oscillators comes down to determining the set of pair synchronization fractional values (in regard to the first oscillator) for $N$-pairs ($m_1/d_1$, $m_2/d_2... m_{N-1}/d_{N-1}$) and to its reduction to a common denominator:

$$(\frac{m_1}{d_1}, \frac{m_2}{d_2}, \frac{m_3}{d_3} \ldots \frac{m_{N-1}}{d_{N-1}}) \rightarrow (\frac{k_2}{k_1}, \frac{k_3}{k_1}, \frac{k_4}{k_1} \ldots \frac{k_N}{k_1}). \qquad (4)$$

For example, a set of pair synchronization for oscillator pairs (№1-№2) and (№1-№3) in Fig. 2a looks like (2/1, 4/3), after reduction to a common denominator (4) we get (2/1, 4/3)→ (6/3, 4/3) and $k_1:k_2:k_3$=3:6:4.

It also should be noted that the algorithm of pair synchronization definition is based on the search of current oscillation peaks $I_{sw}$ synchronous in time [11], in which case we used the threshold value of synchronization efficiency $\eta_{th}$=90%, and the maximum value of $k$ did not exceed $k_{max}$ ≤100.

## 3. Results

The results of synchronization areas modeling for a two-oscillator scheme (see Fig. 1a) are given in Fig. 4. Control parameters are oscillator feed currents $I_{p1}$, $I_{p2}$, noise and coupling strength values are $U_{in}$=20mV and Δ=0.2V. It can be seen that there is a whole family of synchronization areas that are called Arnold tongues [9]. The number of possible variants of synchronous states $N_s$ (information capacity of associative memory) in which the system may exist while the control parameters are varied, is $N_s$ =9. Based on the method presented in section 2.1, to plot vectors E it is necessary to set the neural network at such initial point of control parameters (S) that is located outside the system's synchronization area. At non-synchronous state a system may potentially reach one of the various synchronization states $N_s$, the transitions are denoted as vectors $\mathbf{E}_1...\mathbf{E}_{NS}$ in the figure. The dimension of the stored vectors in this case is 2, and the coordinates determine current shifts $\mathbf{E}_1$=($\delta I_{p1}$, $\delta I_{p2}$) with respect to the initial point S.

The problem here is that the synchronization areas are long-ranged (of Arnold's tongues shape) therefore there is a wide range of stored patterns coordinates which bring the system into a certain synchronous state. The solution lies in narrowing the dispersion of stored pattern coordinates by using vectors $\mathbf{E'}_1$=($\delta I_1$, $\delta I_2…\delta I_{N-1}$) of dimension less by one than the number of oscillators (N-1), in this case $\mathbf{E'}_1$=($\delta I_1$). In practice this means that we fix the current for one oscillator and vary the currents for the others (see Fig.4, dashed line $I_{p2}$=const). Thus we eliminate the ambiguity of synchronization definition by one of vector coordinates and the areas of possible synchronization are narrowed.

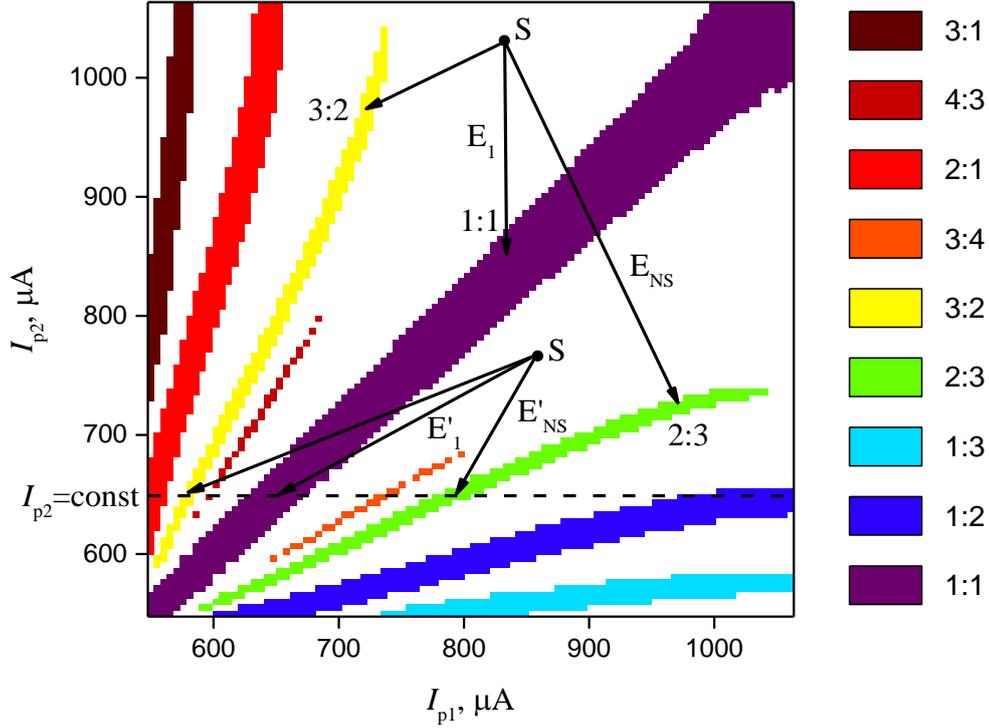

Fig. 4. Example of synchronization areas for a two-oscillator scheme. The arrows show sampled vectors **E** and **E'** in regard to the initial point S.

Fig. 5 shows cross-sections of synchronization areas for a system consisting of three oscillators ("star" and "all-to-all", see Fig. 1 b-c) at fixed current on the first oscillator $I_{p1}$=const, and parameters $U_{in}$=20 mV and $\Delta$=0.2 V that are similar to a two-oscillator scheme.

It can be seen that the synchronization areas are separate isolated regions that are suitable for setting vectors **E'** of dimension 2. In this case, with all other things being equal, the capacity of associative memory $N_s$ depends on the topology and is $N_s$=28 for a "star" connection and $N_s$ =21 for an "all-to-all" connection. The area shape also depends on the topology.

When comparing the values for two- and three-oscillator schemes with the same parameters including the topology we may propose a general rule stating that with the increase of the number of interacting oscillators the information capacity of associative memory $N_s$ increases. Yet, this is evident as the number of freedom degree increases at determining the synchronization value $k_1:k_2:k_3:..k_N$.

Nevertheless, as we show below, at certain parameters there are some exceptions from the general rule.

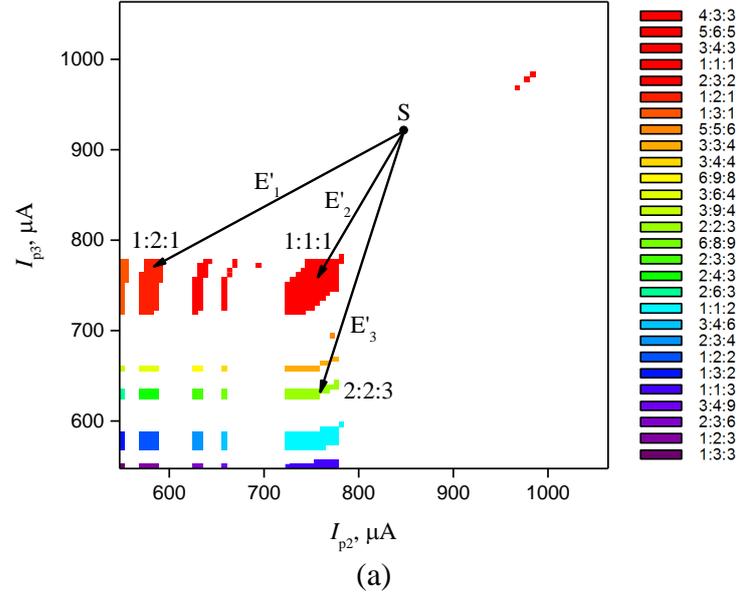

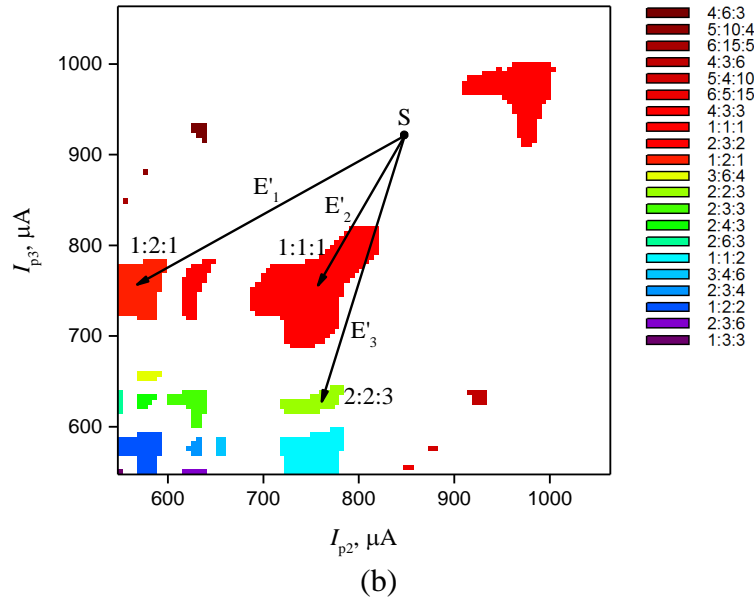

Fig. 5. Example of synchronization areas for three-oscillator schemes "star" (a) and "all-to-all" (b). The arrows show selective vectors **E'** in regard to the initial point S.

Fig. 6 shows the curve $N_s$ versus $\Delta$ at three different configurations of a neural network at the constant noise level $U_{in}$=20 mV. The existence of the main maximum $N_s$ is evident at some optimal value $\Delta_{opt}$, in this case this value is roughly the same as $\Delta_{opt}$ ~0.1 V for all configurations and does not depend on the oscillator number. The existence of the curve maximum $N_s(\Delta)$ reduplicates our results obtained in [10] for a two-oscillator scheme. Inalterability of $\Delta_{opt}$ for different $N$, with all other parameters being equal, might be explained by the fact that with the increase of the number of freedom degree for synchronization order $k_1$:$k_2$:$k_3$:..$k_N$ the value of $N_s$ has the tendency to grow.

We also should note the general tendency for $N_s$ to decrease when the coupling strength $\Delta$ grows above $\Delta_{opt}$, and at its large values the system tends to the lowest possible $N_s=1$ with synchronization value 1:1:1. It is related to the fact that with the increase of $\Delta$ the surface of certain synchronization areas increases. Neighboring areas merge; in this case the synchronization order of the resulting area predominantly consists of lower harmonic numbers. As the dimension of the control parameters is limited such growth of synchronization area surfaces irrevocably results in decrease of their number and value of $N_s$. The insertions in Fig. 6 show the evolution of synchronization areas and demonstrate the effect of their merging with $\Delta$ growth.

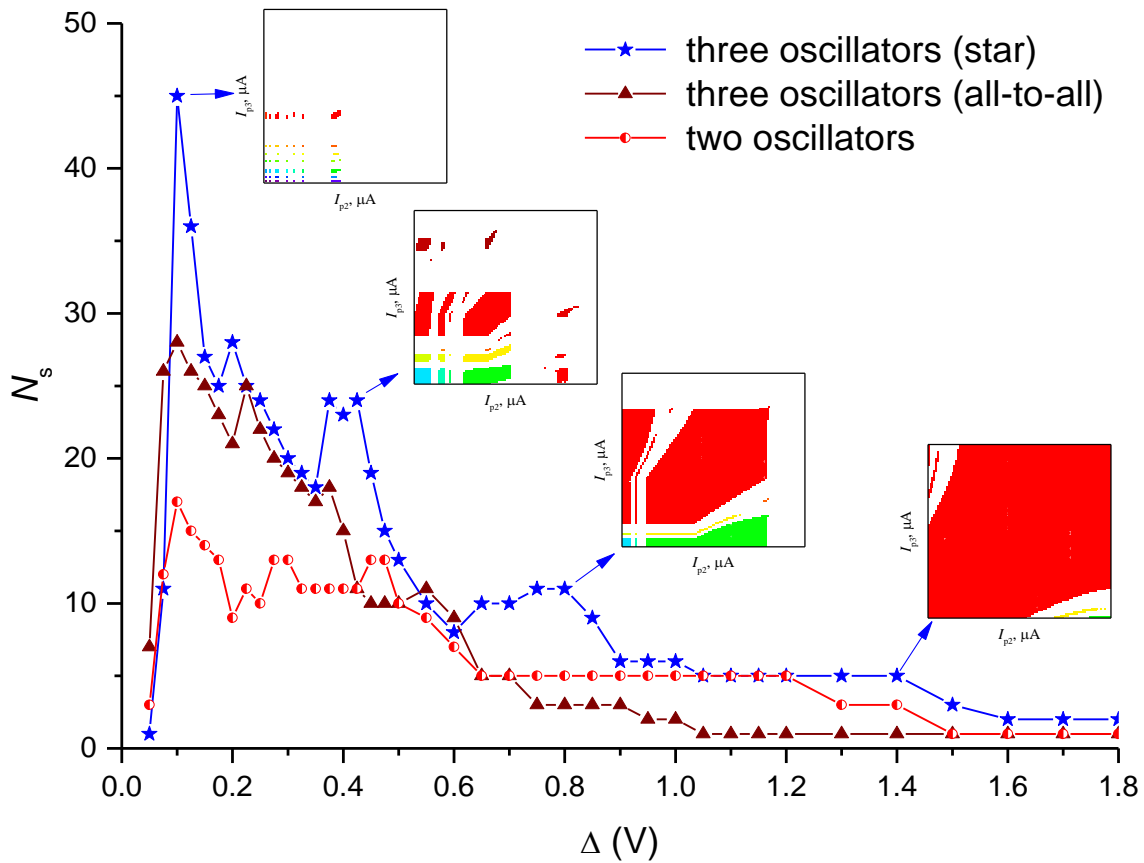

Fig. 6. Curves of associative memory capacity $N_s$ vs. coupling strength value between oscillators $\Delta$ at various configurations of oscillator neural network and constant noise level $U_{in}=20$ mV. The insertions show the evolution of synchronization areas and demonstrate the effect of their merging with $\Delta$ growth.

Besides we should note the existence of local maximums at $\Delta>\Delta_{opt}$. In its turn this is related to the fact that in the presence of noise in the neural network the increase of $\Delta$ may result in development of new synchronization areas at the control parameters values that previously corresponded to the non-synchronous state of the system. Therefore in a general case the curve $N_s(\Delta)$ may have a complicated shape with several maximums as we can see in Fig. 6.

In addition, the initial sharp growth of all three curves at the plot should be noted when $\Delta$ increases from 0 to $\Delta_{opt}$. The latter is due to synchronization effect degradation at $\Delta \rightarrow 0$.

When comparing the curves it may be noted that the increase of the oscillator number in the network leads to the increase of the maximum value of $N_s$ in the system, this does not contradict the rule suggested above. For example, $N_{s\_max}=17$ is for two oscillators, for three-oscillator schemes $N_{s\_max}=28$ and $N_{s\_max}=45$ are for the "all-to-all" and "star" schemes, respectively. At the same time at certain values of coupling strength (for example, at $\Delta=1.1$ V) the value of $N_s$ for a two-oscillator scheme may be even higher. Also the regularity that the "star" topology has higher $N_s$ than the topology of "all-to-all" is observed. All of these things mean that the increase of the coupling number may contribute to the effect of the system desynchronization and decrease of $N_s$, it seems that oscillators prevent each other from synchronization.

Fig. 7 shows the curve of $N_s$ vs. noise level in the system $U_{in}$ at the same coupling strength $\Delta=0.2$ V for three configurations of an oscillator neural network. The general trend for the decrease of $N_s$ at the noise amplitude increase is due to the decrease of the surface of synchronous areas which eventually disappear (see the insertions in Fig. 7).

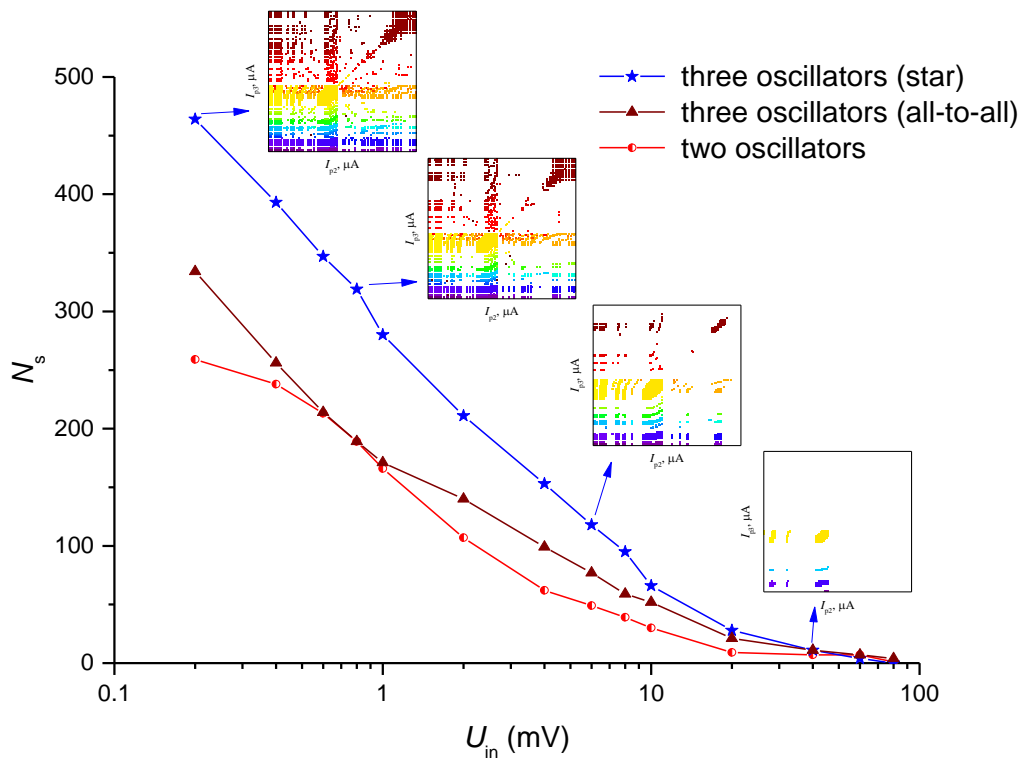

Fig. 7. Curve of associative memory capacity $N_s$ vs. noise level $U_{in}$ at various configurations of an oscillator neural network at the same coupling strength $\Delta = 0.2$ V.

It also should be noted that the general rule is observed; stating that the value of $N_s$ for a tree-oscillator configuration "star" is higher than that for a two-oscillator one. The shapes of curves $N_s(U_{in})$ are similar which indicated that the physics of noise effect on the network is similar and does not depend on the number of oscillators.

Comparing the above curves $N_s(\Delta)$ and $N_s(U_{in})$, it may be seen that the maximum value of associative memory capacity $N_{s\_max}$ in our models may reach $N_{s\_max}$ ~450, at optimal coupling strength value $\Delta=\Delta_{opt}$ and lowered noise $U_{in}=10\mu V$ the capacity increases to $N_{s\_max}$ ~ 650.

We have assumed that the upper limit for $N_{s\_max}$ is in proportion to the exponential function:

$$N_{s\_max} \sim (k_{max})^N, \quad (5)$$

where $k_{max}$ is the maximal number of the harmonic participating in synchronization; and $N$ is the number of oscillators. For example, Fig 8 shows the calculated curve of $N_{s\_max}$ for two coupled oscillators ($N=2$), calculated according to the map of possible synchronization states distribution (Fig. 8b) and its approximation by $N_{s\_max}=0.6\cdot(k_{max})^2$. Here the red squares visualize the allowed states of synchronization $k_1:k_2$, whose number is within the area ($k_1 \leq k_{max}$ and $k_2 \leq k_{max}$, see the example for $k_{max}=16$ in Fig. 8b) and corresponds to $N_{s\_max}$. The white squares show the states that are not included into the calculation as they form simple proportions with one of the states (for example, all states when $k_1=k_2$ is the analogue the state 1:1, and states 2:4 and 4:8 are the analogues to the states 1:2 and so on.)

Fig 8a shows the calculated curve $N_{s\_max}(k_{max})$ for a three-oscillator scheme and its approximation by $N_{s\_max}=0.85\cdot(k_{max})^3$. It can be seen that at the same $k_{max}$ the number of synchronization variants for $N=3$ is much larger than for $N=2$, which indicates the increase of the states number with the increase of the number of freedom degree for synchronization order and is observed in the shown above curves $N_s(\Delta)$ and $N_s(U_{in})$.

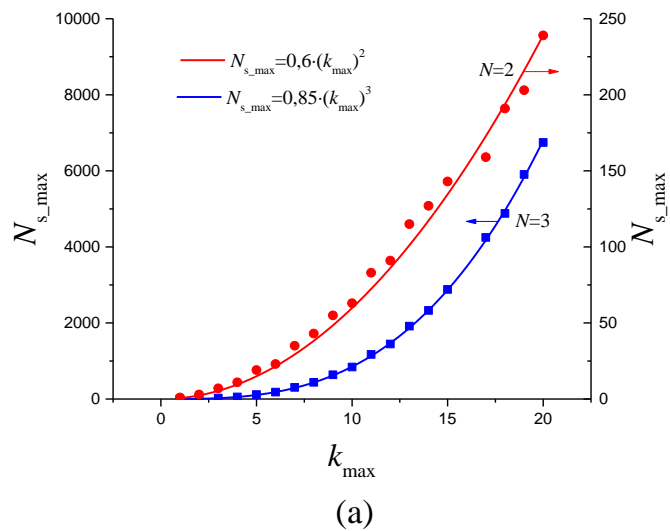

(a)

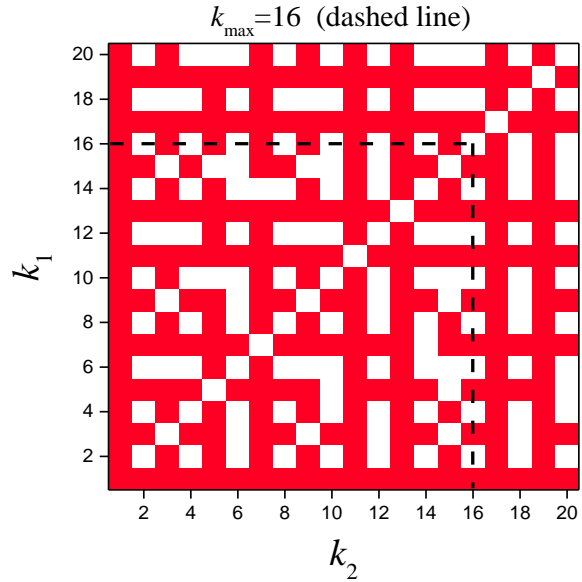

(b)

Fig. 8. a) Curves of maximal information capacity of associative memory $N_{s\_max}$ and their power approximations for two (red curve) and three (blue curve) oscillator schemes vs. the maximum number of harmonic $k_{max}$; b) distribution of allowed states of synchronization (red squares) and forbidden ones (white squares) for $N=2$, the dashed line restricts the area for $k_{max}=16$.

Below we give the example of pattern storage for the scheme "star" at $\Delta=0.2$ V and $U_{in}=20$ mV (see Fig. 5a).

Let us assume that we have to store three patterns with coordinates $\mathbf{E'}_1 = (-264$ μA, $-151$ μA), $\mathbf{E'}_2 = (-91$ μA, $-163$ μA) and $\mathbf{E'}_3 = (-88$ μA, $-290$ μA). Than we should set the initial point S at the coordinate $S = (847$ μA, $921$ μA), where synchronous state does not exist. The number of coordinates in each pattern is one less than the number of oscillators and equals 2, this is necessary to narrow the area of possible synchronization and to lower the recognition ambiguity, as we have explained earlier.

When test vector $\mathbf{T}=\mathbf{E'}_1$ is supplied the neural network is transformed into a synchronous state with the synchronization order 1:2:1. When test vector $\mathbf{T}=\mathbf{E'}_2$ is supplied the neural network is transformed into a synchronous state with the synchronization order 1:1:1. When test vector $\mathbf{T}=\mathbf{E'}_3$ is supplied the neural network is transformed into a synchronous state with the synchronization order 2:2:3.

When the coordinates of test vector $\mathbf{T}$ deviate from the values of the stored patterns $\mathbf{E'}$, within ~1 μA at least, vector $\mathbf{T}$ is also positively recognized.

Thus, we have performed storage and recognition of three various patterns, although the capacity of this system is considerably higher and enables storing up to 28 patterns simultaneously.

## 4. Conclusion

A new method of pattern storage and recognition in an impulse oscillator neural network based on the high-order synchronization effect is presented using computational modeling of two- and three-oscillator schemes with thermally coupled $VO_2$-switches. In comparison with the proposed earlier single-frequency FSK method [8] this method provides significantly higher information capacity of associative memory states $N_s$, where each state of the system is characterized by synchronization order $k_1:k_2:k_3:..k_N$.

A general rule is suggested stating that $N_s$ increases with the increase of the number of interacting oscillators. The modeling demonstrates achievement of $N_s$ of several orders: $N_s$~650 for a three-oscillator scheme and $N_s$~260 for a two-oscillator scheme. It is shown that the theoretical limit for $N_s$ depends on the number of oscillators and can be estimated as an exponential function.

Several regularities of functional (capacitive) characteristics of such ONN have been obtained; in particular, the existence of an optimal coupling strength between oscillators has been revealed, when the number of synchronous states is maximal. A general tendency for the information capacity decrease with the increase of coupling strength and switches inner noise amplitude is also shown.

An algorithm of patterns $\mathbf{E}_1...\mathbf{E}_{NS}$ storage and test pattern $\mathbf{T}$ recognition is proposed. In this algorithm it is better to use the number of coordinates one less than the number of oscillators ($N$-1) that is necessary to narrow the area of possible synchronization and to lower the recognition ambiguity.

Although the research has been performed on a certain model object ($VO_2$ thermally coupled relaxation oscillators), the demonstrated method of associative memory realization is sufficiently general and the fundamental character of the obtained regularities may be the subject of further research of ONN of various mechanisms and oscillators coupling topology.

**Acknowledgments**

This research was supported by Russian Science Foundation (grant no. 16-19-00135).

3, pp. 447–456, 2018.